\documentclass{biometrika}

\usepackage{amsmath}
\usepackage{graphicx}

\usepackage{newtxtext}
\usepackage[subscriptcorrection]{newtxmath}
\usepackage{caption}
\def\Bka{{\it Biometrika}}

\def\BF{\mathrm{BF}}
\newcommand{\N}{\mathcal{N}}

\begin{document}

\jname{Biometrika}
\jyear{2025}
\jvol{112}
\jnum{1}
\cyear{2025}


\markboth{M. M. Lovric}{The Bayes factor reversal paradox}

\title{The Bayes factor reversal paradox}

\author{MIODRAG M. LOVRIC}
\affil{Department of Mathematics and Statistics, Radford University,\\
Radford, Virginia 24142, U.S.A. \email{mlovric@radford.edu}}

\maketitle

\begin{abstract}
In 1957, Lindley published `A statistical paradox' in this journal, revealing 
a fundamental conflict between frequentist and Bayesian inference as sample 
size approaches infinity. We present a new paradox of a different kind: a 
conflict within Bayesian inference itself. In the normal model with known 
variance, we prove that for any two-sided statistically significant result 
at the 0.05 level there exist prior variances such that the Bayes factor 
indicates evidence for the alternative with one choice while indicating 
evidence for the null with another. Thus, the same data, testing the same 
hypothesis, can yield opposite conclusions depending solely on prior choice. 
This answers Robert's 2016 call to investigate the impact of the prior scale 
on Bayes factors and formalises his concern that this choice involves 
arbitrariness to a high degree. Unlike the Jeffreys--Lindley paradox, which 
requires sample size approaching infinity, the paradox we identify occurs 
with realistic sample sizes.
\end{abstract}


\section{Introduction}

The Bayes factor has achieved a position of remarkable prominence in 
statistical methodology. According to Berger (2006, p.\ 378), Bayes factors 
are the `primary tool used in Bayesian inference for hypothesis testing and 
model selection.' Kass \& Raftery (1995) describe the Bayes factor as `a 
summary of the evidence provided by the data in favour of one scientific 
theory over another,' while Goodman (1999) asserts that `the minimum Bayes 
factor is objective and can be used in lieu of the $P$ value as a measure 
of the evidential strength.' These influential endorsements have contributed 
to a growing movement advocating Bayesian hypothesis testing as a superior 
alternative to classical significance testing, with Bayes factors now 
implemented in widely-used software such as JASP (Wagenmakers et al., 2018) 
and the BayesFactor package for R (Rouder et al., 2009). Indeed, Berger \& 
Sellke (1987) argued influentially that $p$-values are `irreconcilable' with 
evidence and should be replaced by Bayes factors, which purportedly provide 
more coherent and interpretable summaries of what the data say.

Not all statisticians share this enthusiasm. In 2016, Christian Robert 
published `The expected demise of the Bayes factor', raising fundamental 
objections to default Bayes factor methodology. Regarding the choice of 
prior scale parameter, Robert observed that `this choice only proceeds from 
an imperative of fat tails in the prior, without in the least solving the 
calibration of the Cauchy scale, which has no particular reason to be set 
to 1. The choice thus involves arbitrariness to a rather high degree.' He 
then issued an explicit call for research: `Given the now-available modern 
computing tools, it would be nice to see the impact of this scale $\gamma$ 
on the numerical value of the Bayes factor.' This paper answers Robert's 
call, and the answer is perhaps more far-reaching than anticipated. We prove 
that in this model the impact of the prior scale is not merely 
quantitative---affecting the magnitude of the Bayes factor---but qualitative: 
it can reverse the scientific conclusion entirely. For any two-sided 
statistically significant result in the normal model with known variance, 
we demonstrate that prior variances exist on both sides of a critical 
threshold, such that one Bayesian concludes in favour of the alternative 
hypothesis while another, analysing the identical data with a different 
prior, concludes in favour of the null. We term this the BFR (Bayes factor 
reversal) paradox.

\section{Mathematical framework}

Following the canonical setup of Lindley (1957), Berger \& Sellke (1987), 
and Lovric (2020), we consider the normal model with known variance, 
which serves as the foundational example for understanding conflicts between 
frequentist and Bayesian inference. Let $X_1, \ldots, X_n$ be a random sample 
from $\N(\mu, \sigma^2)$ with known $\sigma^2 = 1$, and consider testing 
$H_0: \mu = 0$ versus $H_1: \mu \neq 0$. Under Jeffreys' mixed prior 
framework (Jeffreys, 1961), we assign prior probability $\pi_0 = 1/2$ 
to the point null hypothesis $H_0$, and conditional on $H_1$, place a normal 
prior on $\mu$ centred at zero with variance $\tau^2 > 0$, so that 
$\mu \mid H_1 \sim \N(0, \tau^2)$. The parameter $\tau^2$ represents the 
prior variance under the alternative and encodes the researcher's belief 
about plausible values of $\mu$; small values of $\tau$ reflect 
skepticism that $\mu$ is far from zero, while large values express openness to 
substantial departures from the null.

Standard calculations yield the Bayes factor in favour of $H_0$. Letting 
$z = \sqrt{n}\bar{x}$ denote the $z$-statistic, we obtain
\begin{equation}
\label{eq:bf}
\BF_{01}(\bar{x}; \tau^2) = \sqrt{1 + n\tau^2} \cdot 
\exp\left\{-\frac{z^2}{2} \cdot \frac{n\tau^2}{1 + n\tau^2}\right\}.
\end{equation}
This expression reveals the fundamental tension at the heart of Bayesian 
hypothesis testing: the Bayes factor depends not only on the data through 
$z$ but also on the prior specification through $\tau^2$. To study this 
dependence systematically, it proves convenient to reparametrise in terms 
of $k = n\tau^2$, which combines sample size and prior variance into a 
single quantity representing the effective prior precision relative to 
the data. Under this reparametrisation, the Bayes factor becomes
\begin{equation}
\label{eq:bf_k}
\BF_{01}(z; k) = \sqrt{1+k} \cdot \exp\left\{-\frac{z^2 k}{2(1+k)}\right\},
\end{equation}
a form that will facilitate our theoretical analysis.

\section{The flip point and its consequences}

Our analysis begins by examining the behaviour of $\BF_{01}(z; k)$ as a 
function of $k$ for fixed $z$. At the boundary when $k = 0$, corresponding 
to a degenerate prior concentrated at zero, direct substitution yields 
$\BF_{01}(z; 0) = 1$, indicating perfect neutrality between the hypotheses. 
At the opposite extreme, as $k \to \infty$, the Bayes factor grows without 
bound for any fixed $z$, a manifestation of the well-known Jeffreys--Lindley 
effect: with a sufficiently diffuse prior, any fixed dataset will favour 
the null hypothesis. Between these extremes, the behaviour of the Bayes 
factor depends critically on whether the observed $z$-statistic exceeds 
unity in absolute value.

To characterise this intermediate behaviour, we examine the derivative of 
the log Bayes factor at the origin. Straightforward calculus shows that
\[
\frac{d(\ln \BF_{01})}{dk}\bigg|_{k=0} = \frac{1 - z^2}{2},
\]
which is negative when $|z| > 1$. This means that for moderately large 
$z$-statistics, the Bayes factor initially \emph{decreases} as the prior 
variance increases from zero, providing evidence against the null hypothesis. 
Combined with the fact that $\BF_{01} \to \infty$ as $k \to \infty$, this 
implies by continuity that there must exist an intermediate value of $k$ 
at which the Bayes factor equals unity. We call this critical value the 
\emph{flip point} and denote it by $k^*$.

\begin{theorem}
\label{thm:existence}
For $|z| > 1$, there exists a unique flip point $k^* > 0$ satisfying 
$\BF_{01}(z; k^*) = 1$. Moreover, $k^*$ is characterised by the equation
\begin{equation}
\label{eq:flip}
(1+k^*)\ln(1+k^*) = z^2 k^*.
\end{equation}
For $0 < k < k^*$, we have $\BF_{01}(z; k) < 1$, indicating evidence for 
$H_1$; for $k > k^*$, we have $\BF_{01}(z; k) > 1$, indicating evidence 
for $H_0$.
\end{theorem}

The proof proceeds by observing that the derivative of $\ln \BF_{01}$ 
vanishes when $k = z^2 - 1$, establishing that the Bayes factor achieves 
a unique minimum at this point. Since $\BF_{01}$ decreases on 
$(0, z^2 - 1)$ and increases on $(z^2 - 1, \infty)$, with boundary values 
$\BF_{01}(0) = 1$ and $\BF_{01}(\infty) = \infty$, there is exactly one 
solution to $\BF_{01}(k) = 1$ in the region $k > z^2 - 1$. Setting the 
expression in \eqref{eq:bf_k} equal to unity and taking logarithms yields 
the characterizing equation \eqref{eq:flip}. The complete proof appears 
in the Appendix.

The flip point equation admits a useful interpretation. Define the function 
$\phi: (0, \infty) \to (1, \infty)$ by $\phi(k) = (1+k)\ln(1+k)/k$. This 
function is strictly increasing with $\phi(0^+) = 1$ and 
$\phi(\infty) = \infty$, establishing a bijection between $(0, \infty)$ 
and $(1, \infty)$. The flip point is then given by $k^* = \phi^{-1}(z^2)$, 
which shows that larger $z$-statistics correspond to larger flip points: 
more extreme data require more diffuse priors to push the Bayes factor 
above unity. The transcendental equation \eqref{eq:flip} can also be 
expressed in terms of the Lambert $W$ function as 
$k^* = \exp\{W(-z^2 e^{-z^2}) + z^2\} - 1$, though for practical 
purposes numerical solution is straightforward.

\section{The paradox}

The existence of the flip point has immediate implications for Bayesian 
hypothesis testing. Consider any two-sided statistically significant result 
at the conventional $\alpha = 0.05$ level, which requires $|z| > 1.96$. Since 
$1.96 > 1$, Theorem~\ref{thm:existence} guarantees the existence of a 
unique flip point $k^*$ for such data. This leads to our main result.

\begin{theorem}[The BFR paradox]
\label{thm:paradox}
For any two-sided statistically significant result at level $\alpha = 0.05$ 
in the normal model with known variance, there exist prior variances 
$\tau_1^2 < \tau^{*2} < \tau_2^2$ such that:
(i) a Bayesian using prior variance $\tau_1^2$ obtains $\BF_{01} < 1$ 
and concludes in favour of $H_1$;
(ii) a Bayesian using prior variance $\tau_2^2$ obtains $\BF_{01} > 1$ 
and concludes in favour of $H_0$.
Both conclusions are derived from identical data testing identical hypotheses.
\end{theorem}

The result is mathematically universal in this setting: for each 
$z$ with $|z| > 1$, there is a unique flip point $k^*(z)$, so that priors 
with $n\tau^2$ on either side of $k^*(z)$ lead to opposite Bayes factor 
conclusions. In particular, as $|z|$ increases, $k^*(z)$ increases, so 
more extreme $z$-statistics require more diffuse priors to push the Bayes 
factor above unity. Table~\ref{tab:flip} illustrates the flip points for 
various $z$-values. 

For concreteness, consider a study yielding $z = 2.0$ 
with corresponding $p = 0.046$ based on $n = 50$ observations. The flip 
point occurs at $k^* \approx 49.44$, corresponding to a critical prior 
standard deviation of $\tau^* = \sqrt{k^*/n} \approx 0.99$. A Bayesian who 
chooses a moderately concentrated prior $\tau = 0.8$ computes 
$\BF_{01} \approx 0.83$ and concludes that the evidence favours the 
alternative hypothesis. A colleague who chooses the more diffuse 
prior $\tau = 1.5$ computes $\BF_{01} \approx 1.47$ and reaches the opposite 
conclusion: the evidence favours the null. Both priors place most prior 
mass on values of $\mu$ within scientifically plausible ranges, yet the 
conclusions are contradictory.

\begin{figure}[t]
\centering
\includegraphics[width=\textwidth]{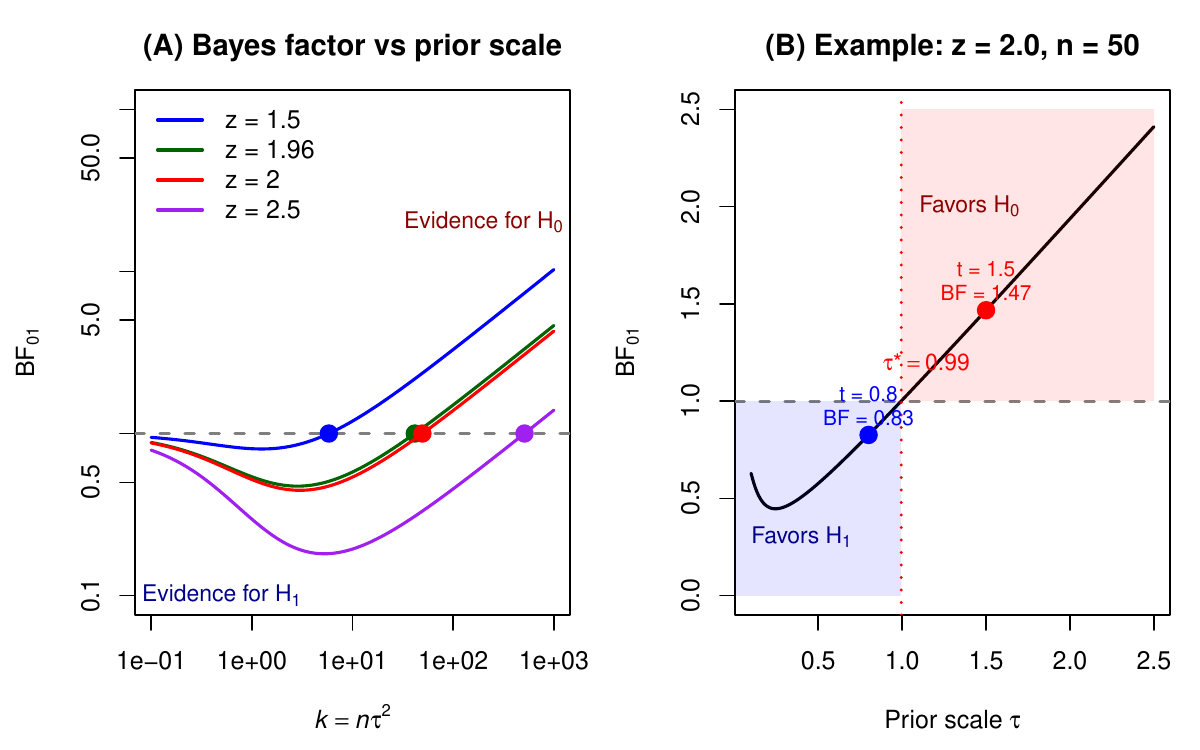}
\caption{The Bayes factor reversal paradox. (A) Bayes factor $\BF_{01}$ as a function of prior scale $k = n\tau^2$ for different $z$-values. The horizontal dashed line marks $\BF_{01} = 1$; dots indicate flip points $k^*$ where evidence reverses direction. Below the line, evidence favours $H_1$; above, it favours $H_0$. (B) Detailed view for $z = 2.0$, $n = 50$. Two analysts using priors $\tau = 0.8$ and $\tau = 1.5$---both scientifically reasonable---reach opposite conclusions from identical data: $\BF_{01} = 0.83$ (favours $H_1$) versus $\BF_{01} = 1.47$ (favours $H_0$). The vertical dashed line marks the critical prior scale $\tau^* = 0.99$.}
\label{fig:bfr}
\end{figure}

Figure~\ref{fig:bfr} illustrates this reversal graphically. Panel~(A) shows how the Bayes factor varies with prior scale for several $z$-values, with flip points marked where evidence changes direction. Panel~(B) demonstrates the paradox concretely: two reasonable prior choices yield opposing conclusions from identical data.

The paradox becomes even more striking in large studies. Consider a 
clinical trial with $n = 5000$ that yields $z = 1.96$ (exactly $p = 0.05$), 
corresponding to a sample mean of $\bar{x} = 0.028$. Here the flip point 
occurs at $\tau^* \approx 0.09$, an extremely concentrated prior. 
Consequently, almost any standard prior choice falls above the flip 
point: the JZS default ($\tau \approx 0.707$) yields $\BF_{01} \approx 7.3$, 
indicating moderate evidence \emph{for the null hypothesis} despite 
the statistically significant $p$-value. A diffuse prior ($\tau = 1.0$) 
gives $\BF_{01} \approx 10.4$, strong evidence for the null, while a 
very diffuse prior ($\tau = 2.0$) gives $\BF_{01} \approx 20.7$. Only an 
unusually skeptical analyst with $\tau = 0.05$ obtains $\BF_{01} \approx 0.62$, 
weak evidence for the alternative. This represents a 33-fold swing in 
the Bayes factor across scientifically defensible prior choices.

\begin{table}
\tbl{Flip points $k^*$ for various $z$-values and corresponding critical 
prior standard deviations $\tau^* = \sqrt{k^*/n}$}
{\begin{tabular}{cccccc}
$z$ & $z^2$ & $p$-value & $k^*$ & $\tau^*$ ($n=50$) & $\tau^*$ ($n=100$) \\
1.50 & 2.25 & 0.134 & 5.82 & 0.34 & 0.24 \\
1.96 & 3.84 & 0.050 & 41.58 & 0.91 & 0.64 \\
2.00 & 4.00 & 0.046 & 49.44 & 0.99 & 0.70 \\
2.50 & 6.25 & 0.012 & 510.72 & 3.20 & 2.26 \\
3.00 & 9.00 & 0.003 & 8093.08 & 12.72 & 9.00
\end{tabular}}
\label{tab:flip}
\end{table}

\section{Discussion}

The paradox we have identified differs fundamentally from previously known 
conflicts in statistical inference. The Jeffreys--Lindley paradox (Lindley, 
1957) demonstrates a disagreement between frequentist and Bayesian paradigms, 
but this disagreement manifests only asymptotically as sample size approaches 
infinity and arguably reflects a genuine philosophical difference about the 
nature of evidence. Lovric (2020) showed that Bayesian credible intervals 
and Bayes factors can disagree for finite samples. Related phenomena have 
been discussed under the heading of prior--data conflict (Evans \& Moshonov, 
2006; Nott et al., 2020); the contribution here is the closed-form flip-point 
equation, which allows exact calibration of the critical threshold. The 
present paradox reveals a conflict within Bayesian inference itself: two 
analysts who share the same likelihood, the same prior model structure, and 
even the same commitment to Bayesian principles can reach opposite conclusions 
based solely on their choice of a scale parameter. This conflict occurs with 
realistic sample sizes and across a wide range of statistically 
significant results, making it directly relevant to applied statistical 
practice.

Our findings call into question certain strong claims that have been made 
for Bayes factors as measures of evidence. Kass \& Raftery (1995) describe 
the Bayes factor as `a summary of the evidence provided by the data,' but 
our results show that the Bayes factor summarises something more than the 
data alone---it inextricably confounds data with prior specification. 
Goodman (1999) claimed that Bayes factors are `objective' and possess `a 
sound theoretical foundation,' yet any claim to objectivity is undermined 
when different reasonable prior scales yield not merely different magnitudes 
of evidence but opposite conclusions about the direction of the evidence. 
The widely-used `default' priors of Rouder et al.\ (2009), implemented in 
JASP and the BayesFactor package with a Cauchy scale of $r = \sqrt{2}/2 
\approx 0.707$, represent one particular choice among many that may fall 
on either side of the flip point depending on the data.

Although our analysis uses a normal prior on $\mu$ under $H_1$ in order to 
obtain the closed-form expression \eqref{eq:bf_k}, numerical experiments 
with heavy-tailed priors such as the Cauchy distributions underlying the 
JZS $t$-tests of Rouder et al.\ (2009) exhibit analogous flip behaviour: 
for fixed test statistics and sample sizes, varying the prior scale can 
move the Bayes factor across unity. For example, with $z = 2.0$ and $n = 50$, 
a Cauchy scale of $r = 0.6$ yields $\BF_{01} \approx 0.9$ (favouring $H_1$), 
while $r = 1.0$ yields $\BF_{01} \approx 1.3$ (favouring $H_0$); notably, 
the default JZS scale $r = \sqrt{2}/2 \approx 0.707$ produces 
$\BF_{01} \approx 1.00$, placing it almost exactly at the flip point for 
this dataset. The normal model considered here thus serves as a tractable 
prototype illustrating a phenomenon that appears to be general.

The relationship between our results and the critique of $p$-values deserves 
comment. Berger \& Sellke (1987) argued that $p$-values are `irreconcilable' 
with evidence and should be replaced by Bayes factors. Our results suggest 
a complementary perspective. While $p$-values are uniquely determined by 
the data and the null hypothesis---all frequentists will compute the same 
$p$-value from the same data---Bayes factors depend on an additional 
specification, the prior scale. With $p$-values, statisticians may disagree 
about interpretation, but they at least agree on the number; with Bayes 
factors, different prior choices can lead to disagreement about whether 
the evidence favours the null or the alternative.

These results have implications for the objective Bayes programme. 
Berger (2006) stated that `a major goal of statistics, indeed science, is 
to find a completely coherent objective Bayesian methodology for learning 
from data.' Our analysis suggests that achieving full objectivity in 
hypothesis testing with continuous priors on $\mu$ faces 
inherent difficulties. There is no choice of $\tau^2$ that avoids the 
existence of the flip point; the flip point is a mathematical consequence 
of the Bayes factor structure itself. Different analysts with different 
but defensible priors may reach opposite conclusions. Under zero--one 
loss, the posterior median decision also flips at the same $k^*$, showing 
that the decision---not merely the evidence summary---is prior-dependent.

Robert's 2016 conjecture that the choice of prior scale `involves 
arbitrariness to a rather high degree' is thus formalised and made precise. 
The arbitrariness is not merely aesthetic or philosophical; it can determine 
the scientific conclusion itself. One might argue that this is simply 
prior sensitivity, a known feature of Bayesian inference. However, the 
BFR paradox demonstrates something more troubling: not merely that the 
\emph{strength} of evidence varies with the prior, but that its 
\emph{direction} reverses. A Bayes factor of $0.5$ and one of $2.0$ do not 
represent different strengths of evidence for the same hypothesis---they 
represent evidence for opposite hypotheses. When the choice of a scale 
parameter can reverse whether we conclude for or against a hypothesis, 
the methodology provides less objectivity than some of its proponents 
have claimed. This does not mean that Bayesian methods lack value---they 
remain essential tools for incorporating prior information, quantifying 
uncertainty, and updating beliefs coherently---but it does suggest that 
Bayes factors for point null hypotheses may be less suitable as default 
measures of evidence than has sometimes been argued.

Several directions for future research emerge from this work. The present 
analysis focuses on the normal model with known variance, following the 
tradition established by Lindley; extending the results to $t$-tests, 
analysis of variance, and regression settings would establish the generality 
of the phenomenon across common statistical applications. The theoretical 
framework developed here---particularly the flip point equation and its 
properties---may provide useful tools for such extensions. Additionally, 
investigating whether alternative Bayesian approaches to hypothesis testing, 
such as those based on credible intervals or posterior predictive checks, 
are subject to analogous difficulties would help clarify the scope of the 
problem we have identified.

\section*{Declaration of the use of generative AI and AI-assisted technologies}

The author used AI tools (Claude, ChatGPT, MathGPT, Gemini, Grok, Copilot, DeepSeek, Kimi 2) solely for minor editorial assistance, formatting, and sanity-checking selected calculations. All mathematical proofs, theoretical results, and methodological contributions were independently developed and rigorously verified by the author, who takes full responsibility for the manuscript's content.

\section*{Acknowledgement}

The author thanks Christian Robert for his prescient 2016 critique of Bayes 
factors, which inspired this investigation.

\section*{Data and code availability}

The R code that reproduces all numerical results and figures is available 
in the Supplementary Material, which is available at \Bka\ online. Code is 
also available at \texttt{https://github.com/MLovricStats/BFR-Paradox}. 
An interactive demonstration is provided at 
\texttt{https://sites.radford.edu/\~{}mlovric/BFR-Paradox-Interactive-Tool.html}.

\section*{Supplementary material}

Supplementary material available at \Bka\ online includes R code for 
computing flip points, verifying all theoretical results, reproducing 
Table~\ref{tab:flip} and Figure~\ref{fig:bfr}, and demonstrating the 
33-fold swing in the large-sample scenario.

\appendix

\appendixone
\section{Technical proof}

\begin{proof}[of Theorem~\ref{thm:existence}]
We establish the existence and uniqueness of the flip point through 
analysis of the Bayes factor as a function of $k$.

First, observe the boundary behaviour. At $k = 0$, direct substitution 
into \eqref{eq:bf_k} yields $\BF_{01}(z; 0) = \sqrt{1} \cdot \exp(0) = 1$. 
As $k \to \infty$, we have $k/(1+k) \to 1$, so the exponential term 
approaches $\exp(-z^2/2)$, a positive constant, while the $\sqrt{1+k}$ 
term diverges. Hence $\BF_{01}(z; k) \to \infty$.

Next, we analyse the derivative. Taking the logarithm of \eqref{eq:bf_k}:
\[
\ln \BF_{01} = \frac{1}{2}\ln(1+k) - \frac{z^2 k}{2(1+k)}.
\]
Differentiating with respect to $k$:
\[
\frac{d(\ln \BF_{01})}{dk} = \frac{1}{2(1+k)} - \frac{z^2}{2(1+k)^2} 
= \frac{(1+k) - z^2}{2(1+k)^2}.
\]
This derivative equals zero when $1 + k = z^2$, i.e., at $k = z^2 - 1$. 
For $|z| > 1$, this critical point lies in $(0, \infty)$. The second 
derivative is positive at this point, confirming it corresponds to a 
minimum of $\BF_{01}$.

Since $\BF_{01}(0) = 1$ and the function initially decreases (as the 
derivative at $k = 0$ equals $(1 - z^2)/2 < 0$ when $|z| > 1$), the 
minimum value is strictly less than 1. Combined with $\BF_{01}(k) \to \infty$ 
as $k \to \infty$, continuity guarantees exactly one solution to 
$\BF_{01}(k) = 1$ in the region $k > z^2 - 1$.

Finally, setting \eqref{eq:bf_k} equal to 1 and taking logarithms:
\[
\frac{1}{2}\ln(1+k^*) = \frac{z^2 k^*}{2(1+k^*)}.
\]
Multiplying both sides by $2(1+k^*)$ yields the characterizing equation 
$(1+k^*)\ln(1+k^*) = z^2 k^*$.
\end{proof}

\bibliographystyle{biometrika}

\end{document}